\documentclass[a4paper,11pt]{article}

\usepackage{jheppub}
\usepackage{amssymb}
\usepackage{amsfonts}
\usepackage{amsmath}
\usepackage{amsthm}
\usepackage{enumerate}
\usepackage{subcaption}
\usepackage{appendix}
\usepackage{comment}
\usepackage{graphicx}
\usepackage{braket}
\usepackage{scrextend}
\usepackage{bbold}
\usepackage{verbatim}
\usepackage{setspace}

\newcommand{\mpl}[0]{\ensuremath{M_\text{Pl}}}

\newcommand{\ads}[0]{\ensuremath{\text{AdS}_2}}
\newcommand{\stwo}[0]{\ensuremath{\text{S}^2}}
\newcommand{\adss}[0]{\ensuremath{\text{AdS}_2 \times S^2}}

\newcommand{\adsdcftd}[2]{\ensuremath{\text{AdS}_#1/\text{CFT}_#2}}

\title{A Semiclassical, Entropic Proof of a Weak Gravity Conjecture}
\author{Zachary Fisher and Christopher J. Mogni}
\affiliation{Berkeley Center for Theoretical Physics and Department of Physics\\
University of California, Berkeley, CA, 94720-7300, USA\medskip\\ 
Theoretical Physics Group, Lawrence Berkeley National Laboratory\\
Berkeley, CA 94720-8162, USA}
\emailAdd{zach@zkf.me}
\emailAdd{cmogni1@berkeley.edu}

\abstract{We present a semiclassical proof of the weak gravity conjecture in $D = 4$ spacetime dimensions for scalar matter gauged under a $U(1)^N$ gauge group. We compute the non-perturbative macroscopic entropy of a scalar field in an extremal black hole background at the level of linearized backreaction on the metric. The scalar field is assumed to violate or saturate the weak gravity conjecture. The scalar contributes a logarithmic correction to the entropy in the black hole geometry that outgrows the classical contribution. We demonstrate that the entropy of the gauged scalar violates the generalized second law in the limit of large black hole charge. Our result suggests that entropy inequalities may directly discriminate between effective field theories that live in the landscape versus the swampland.}

\begin{document}

\maketitle

\section{Introduction}

In its simplest incarnation \cite{ArkMot06}, the weak gravity conjecture states that a consistent, quantized theory of gravity coupled to an Abelian gauge theory must contain at least one charged, massive particle satisfying
\begin{equation}
    m \le q \mpl,
\end{equation}
\noindent where $m$ is the particle mass and $q$ the particle charge. Because Newton's constant $G_N = 1/\mpl^2$, the bound implies gravity is the weakest force. All known string compactifications with Abelian gauge forces satisfy the conjecture. Moreover, it reconciles the absence of global symmetries in string theory with the $q\rightarrow 0$ limit of Abelian gauge theories. Within the context of perturbative string theory, the authors of \cite{Heidenreich:2016aqi} demonstrate that modular invariance of effective worldsheet theories evidently implies a version of the conjecture. Extensions of the weak gravity conjecture apply to $p$-form gauge fields of any $p \ge 0$ in arbitrary spacetime dimensions $D \ge 3$ \cite{ArkMot06}. In this paper, we focus on $p = 1$, $D = 4$.

Although string theory automatically satisfies the weak gravity conjecture, the authors of \cite{ArkMot06} use black holes to argue that all healthy effective field theories should obey a weak gravity conjecture. Suppose a black hole has charge $Q$ and mass $M$. Assuming cosmic censorship, $M \ge Q \mpl$. The black hole may decay via Hawking radiation or Schwinger pair production. For black holes far from extremality, Hawking radiation dominates. If the black hole only emits charged particles with charge $q$, and mass $m$, then conservation of charge implies that $Q/q$ particles are produced. The black hole evolves to a state with mass $m Q/q$, which is less than $M$ by conservation of energy. Through this process, the black hole approaches extremality, $Q/M = 1$.\footnote{We work in Planck units, i.e. $\mpl = 1$.} At extremality, the black hole's temperature is zero, and Hawking radiation ceases. Such a black hole is stable unless there is a charged particle with $q/m > 1$, in which case particle-antiparticle pairs are produced via Schwinger pair production. Pair production emits charged matter from the black hole; the black hole is no longer extremal. On the other hand, if the weak gravity conjecture is violated, a large number of stable extremal black hole states exist in the full quantum theory.\footnote{Previously, it was believed that the presence of a large number of stable, Planck sized extremal black hole states would violate known entropy bounds \cite{Banks:2006mm}. However, Casini \cite{Cas08} casts doubt on this assertion by carefully examining properties of relative entropy, showing that entropy bounds may not necessarily rule out remnants.} While a proliferation of stable quantum states does not itself signal a sickness from the effective field theory's perspective, it does appear physically undesirable.

Recent research directions have focused on sharpening and defining the weak gravity conjecture using effective field theory. The authors of \cite{CheRem14} propose a stronger form of the weak gravity conjecture by studying matter gauged under a $U(1)^N$ symmetry group. They claim that the convex hull of the charge-to-mass vectors $z_i$ for each species $i$ of particles gauged under the $U(1)^N$ group must contain the unit ball $|z_i| \le 1$. The same authors also attempt to frame the conjecture in terms of unitarity and causality of infrared scattering amplitudes \cite{CheRem14b}, but \cite{Har15} discusses counterexamples to their original argument. A series of papers \cite{Heidenreich:2015wga,Heidenreich:2015nta,Heidenreich:2016aqi} combine intuition from black hole physics with considerations from effective field theory to sharpen the conjecture and to cast doubt on the consistency of field theories that violate it, such as large field axion inflationary models.

Nonetheless, an inherent sickness in effective field theories violating the weak gravity conjecture has eluded discovery. Proving the conjecture from a ``bottom-up'' perspective within the realm of flat space effective field theory may prove too difficult, or impossible. Consequently, effective field theories on large black hole backgrounds provide an ideal setting to test the conjecture without needing to invoke assumptions or intuition from some unknown UV theory. Presumably, we should be able to treat the near horizon physics of large black holes semi-classically due to the smallness of the Ricci curvature. One expects that entanglement of macroscopic fields across the horizon should tell us something about the underlying gravitational theory, even in a semi-classical setup.

Let us suppose a proliferation of stable black hole states is a property of sick effective field theories. It is plausible that the sickness would manifest itself by violating known properties of semi-classical entropy. The past decade has seen immense progress in unravelling entropy inequalities that encode deep connections between field theory and semi-classical gravity \cite{Cas08,BouCas14a,BouCas14b,BouFis15a,BouFis15b}. It is natural to speculate that macroscopic entropy might be powerful enough to discriminate between effective field theories that live in the landscape or swampland.

Sen \emph{et al.} laid the foundation to study black hole entropy in effective field theory \cite{Banerjee:2011jp,Bhattacharyya:2012wz,Sen:2011ba,Sen11,Sen:2008vm,Sen:2012dw}. They calculate logarithmic corrections to black hole entropy from the Euclidean path integral over the near horizon black hole geometry. One may work with the near horizon geometry directly because of the attractor mechanism, which Sen \emph{et al.} also show applies to non-BPS black holes in the near-extremal limit. They further justify their methodology by matching the macroscopic entropy results with microscopic state counting using the \adsdcftd{2}{1} duality. In low energy effective theories descending from string theory, the results match on both sides of the duality.

These papers do not address the macroscopic entropy due to fields interacting with the background field strength. The presence of a background electric flux modifies the effective masses of the matter fields near the horizon. The flux depends on the radius of the black hole. If the fields have sufficiently small mass relative to their charge, the coupling to the flux renders the near-horizon geometry unstable. It decays rapidly due to Schwinger pair production of particle-antiparticle pairs, which precludes us from calculating the macroscopic entropy with Sen's formalism. On the contrary, whenever the weak gravity conjecture is violated or saturated, the geometry is stable. No symmetry protects the stability of the extremal black hole in the non-supersymmetric theories we consider. We expect that perturbations of the extremal geometry may alter the black hole entropy in a way incompatible with known entropy inequalities after we account for quantum effects.

The purpose of this paper is to confirm this hypothesis. To our knowledge, this is the first concrete demonstration that entropy inequalities may discriminate between effective theories that live in the swampland or landscape in a controlled, semi-classical environment. We consider $D=4$ scalar matter gauged under a $U(1)^N$ gauge group in a large, extremal black hole background. The scalar matter violates the weak gravity conjecture. The scalar is minimally coupled to the gravitational and gauge fields. We do not include any non-renormalizable interactions or scalar-scalar interactions. We compute the exact, non-perturbative macroscopic contribution of the gauged scalar to the entropy of the black hole.\footnote{We hold the external gauge and gravitational fields fixed. Determining the full macroscopic entropy requires the gauge and gravitational sectors as well. Note however that the quantum corrections of fields neutral under the gauge symmetry are generically subleading. The calculation is exact in the semiclassical limit because the action is quadratic in the gauged scalar field.} We choose a renormalization condition that sets an extremal black hole solution with large charge $|\vec{Q}|$ to its classical value. We consider a perturbation to the black hole whereby a neutral particle with energy $E$ crosses the black hole horizon. We demonstrate that any small perturbation violates the second law for a sufficiently large initial black hole solution.\footnote{What we refer to as the second law is typically referred to as the generalized second law in the literature. We omit the word ``generalized'' because the generalized second law is the second law once one accounts for all sources of entropy.} Consequently, we prove the weak gravity conjecture for a single scalar.

\subsection{Related work}

Qualitatively similar results to our entropy calculation appear in \cite{CotShi16}. However, not all of their quantitative results match ours exactly. We believe that this results from the formalism they use to calculate the entropy of the black hole, which is not exactly equivalent to ours. We also believe that their conclusions and interpretation of results differ significantly enough from our own. Moreover, they do not attempt to prove the weak gravity conjecture using entropy inequalities, although they allude to this possibility.

A separate application of the second law towards understanding the weak gravity conjecture appears in \cite{Hod17}, which appeared during the preparation of this manuscript. However, their calculation is orthogonal to ours. Their paper argues for the weak gravity conjecture using a bound on relaxation rates of quasinormal modes of near-extremal black holes. Although related to the second law, the connection is indirect: the second law implies the relaxation rate bound, which in turn implies the weak gravity conjecture. In this paper, we present a a more direct link between the second law and the weak gravity conjecture.

\section{Setup}

Consider a charged, non-rotating black hole. The metric is
\begin{equation}
  ds^2 = - \frac{(r - r_+)(r - r_-)}{r^2} dt^2 +  \frac{r^2}{(r - r_+)(r - r_-)} dr^2 + r^2 d\Omega_{S^{D-2}}^2,
\end{equation}
\noindent where
\begin{equation}
  r_{\pm} =  M \pm \sqrt{M^2 - |\vec{Q}|^2}
\end{equation}
\noindent are the outer and inner horizons of the black hole in units where $\mpl = 1$. $M$ is the ADM mass of the black hole spacetime. The black hole is a solution of Einstein's equations, where the stress-energy tensor descends from a $U(1)^N$ gauge theory action. The classical action is
\begin{equation}
  S_0 = \frac{1}{16\pi} \int d^D x \sqrt{\operatorname{det} \, g} \, \left ( \mpl^2 R - \sum\limits_{n = 1}^N F_{\mu\nu}^{(n)}F^{(n) \, \mu\nu} \right ).
\end{equation}
\noindent where $g$ is the determinant of the spacetime metric, $R$ is the Ricci scalar, and $F^{(n)}$ is the field strength for the $n^{\text{th}}$ gauge field. The background gauge fields $A_{\mu}^{(n)}$ are a Coulomb potential in the appropriate gauge:
\begin{equation}
  A_\mu^{(n)} = \left(\frac{Q^{(n)}}{r}, 0, \dots, 0\right).
\end{equation}
In the extremal limit, $M \rightarrow |\vec{Q}|$, the coordinates of the horizons degenerate to
\begin{equation}
r_E^2 = |\vec{Q}|^2.
\end{equation}
\noindent We may compute the macroscopic entropy of the classical geometry and quantum fluctuations about it using the near-horizon geometry \cite{Sen11}.\footnote{This is computationally beneficial because there are no conifold singularities in the near-horizon geometry.} After an appropriate choice of coordinates and Wick rotation to Euclidean signature, the near-horizon geometry in $D = 4$ spacetime dimensions is described by\footnote{Roughly, $\cosh \eta$ corresponds to the proper distance from the outer horizon in the near-horizon geometry. Details on deriving this form of the metric by taking the near-horizon and extremal limits may be found in \cite{Sen:2008vm}. The utility of working with this form of the metric is that there are no conical singularities.}
\begin{equation}
  ds^2 = r_E^2 \left ( d\eta^2 + \sinh^2\eta \, d\theta^2 + d\psi^2 + \sin^2\psi \, d\varphi^2 \right ),
\end{equation}
\noindent where $\theta$ is $2\pi$-periodic.\footnote{The coordinate $\theta$ is related to Euclidean time by a rescaling. The Euclidean time coordinate has infinite periodicity for extremal black holes. The normalization of Euclidean time such that it has period $2\pi$ permits us to find a finite result for the macroscopic entropy.} The near-horizon extremal metric factorizes as \adss{}.

The macroscopic entropy of the black hole may be calculated by calculating the effective action for the quantum fluctuations about the classical background. We work with the normalization of the Euclidean action in \cite{Sen11}. The effective action splits into a classical ($S_0$) and quantum ($\Delta W_{\text{eff}}$) component:
\begin{equation}
    W_{\text{eff}} = S_0+ \Delta W_{\text{eff}}.
\end{equation}
\noindent Using
\begin{equation}
    F^{(n)}_{\eta \theta} = Q^{(n)} \operatorname{sinh}\eta
\end{equation}
\noindent and
\begin{equation}
    R = 2/r_E^2,
\end{equation}
\noindent we obtain
\begin{equation}
    S_0 = - 2 \beta r_E -4\pi r_E^2,
\end{equation}
\noindent where $\beta = 4\pi r_E \cosh \eta_0$ is the inverse temperature of the near-extremal black hole induced by the \adss{} boundary cutoff.\footnote{The cutoff is implicitly taken to infinity, indicating that the black hole has a temperature that limits to zero, as expected for near-extremal black holes.} The first term in the classical part of the effective action is the classical entropy. The second is the classical black hole energy multiplied by the inverse temperature of the black hole.

Quantum corrections to the effective action may be calculated by splitting each field $\Phi$ into their classical background value $\Phi_{\text{cl}}$ and fluctuations about the background $\Phi_{\text{q}}$:
\begin{equation}
   \Phi(x) = \Phi_c(x) + \Phi_q(x).
\end{equation}
\noindent If we truncate the action for the fluctuations about the background at quadratic order, we may calculate the one-loop contribution to the effective action. This classical action changes by $\Delta W_{\text{eff}}$ \cite{Sen11}:
\begin{equation}\label{eq:first}
  \Delta W_{\text{eff}} = \int d^4 x \, \sqrt{\operatorname{det} g}\, \Delta \mathcal{L}_{\text{eff}} = \frac{1}{2}\pi r_E^4 \left ( \cosh\eta_0 - 1  \right ) \Delta \mathcal{L}_{\text{eff}},
\end{equation}
\noindent where $\Delta \mathcal{L}_{\text{eff}}$ is the effective Lagrangian. The first term corrects the ground state energy, regularized by an infrared cutoff $\eta_0$.\footnote{This IR cutoff renders the volume of E\ads{} finite.} The second term corrects the macroscopic entropy \cite{Sen11}:
\begin{equation}\label{eq:SequalsL}
  S_{\text{quant}} = -\frac{1}{2}\pi r_E^4 \Delta\mathcal{L}_{\text{eff}}.
\end{equation}
\noindent From this expression, it is explicitly clear that in the near-extremal limit, where we can take $\beta \rightarrow \infty$, that the difference in entropies between two near-extremal geometries automatically satisfies the first law of thermodynamics.

Calculating the quantum correction to the macroscopic entropy reduces to calculating $\Delta \mathcal{L}_{\text{eff}}$.\footnote{Some places in the literature refers to the quantum correction we compute as $S_{\text{out}}$, and the macroscopic entropy as $S_{\text{gen}}$.} The evolution operator along Euclidean worldline time for a particle with worldline Hamiltonian $\hat{H}$ is the heat kernel \cite{Schwartz:2013pla,Vas03}
\begin{equation}
  K(x,x';s) = \langle  x'  |\, e^{-s\hat{H}} \, |  x \rangle.
\end{equation}
\noindent To derive $\hat{H}$ for fluctuations of a scalar field about a classical background, consider the minimally gauged scalar field action:
\begin{equation}
  S_{\phi} = \int d^4 x \sqrt{\operatorname{det} \, g} \, \left ( -g^{\mu\nu} \overline{\phi} \left ( \nabla_{\mu} + q A_{\mu} \right ) \left ( \nabla_{\nu} + q A_{\nu} \right ) \phi + m^2 \overline{\phi}\phi \right ),
\end{equation}
\noindent where $\nabla_{\mu}$ is the covariant derivative compatible with the metric $g_{\mu\nu}$. The worldline Hamiltonian for the $\phi$ field is
\begin{equation}
  \hat{H} = -g^{\mu\nu} \left ( \nabla_{\mu} + q A_{\mu} \right ) \left ( \nabla_{\nu} + q A_{\nu} \right ) + m^2.
\end{equation}
\noindent Inserting $\hat{H}$ into the heat kernel, we obtain the quantum correction to the effective action:
\begin{equation}
  \Delta \mathcal{L}_{\text{eff}} = \frac{1}{2} \int\limits_{\varepsilon}^{\infty} \frac{ds}{s}\! \int d^4 x \sqrt{\operatorname{det} g} \, K(s),
\end{equation}
\noindent where $K(s) \equiv K(x,x;s)$.\footnote{$K(s)$ is independent of $x$ by translational symmetry.} A small distance cutoff $\varepsilon$\footnote{With dimensions length squared.} must be imposed due to divergences at the lower bound of the $s$ integral.

We may calculate the heat kernel in two ways. Perturbatively, we may perform an expansion of the heat kernel for small $s$ \cite{Vas03,Solodukhin11}. We express the heat kernel in powers of the Riemann curvature, field strengths, and their contractions, multiplied by the appropriate power of $s$. The geometric expansion yields the perturbative, one-loop contribution to the effective action. This is the familiar small $s$ expansion of the heat kernel. For an arbitrary scalar field, this expansion reads

To find an \emph{exact} solution, we decompose the heat kernel as a sum of the eigenfunctions $f_n(x)$ and eigenvalues $\kappa_n$ of $\hat{H}$ \cite{Vas03}:
\begin{equation}\label{eq:heatFull}
  K(x,x';s) = \sum\limits_n f_n(x) f_n^*(x) e^{-\kappa_n s}.
\end{equation}
\noindent By performing the sum, we obtain the resummed one-loop contribution to the effective action. If the action is quadratic in the field $\Phi$, then the resummed one-loop correction is the \emph{exact} correction to the effective action for the $\Phi$ field in the presence of \emph{fixed, external} $A_{\mu}^{(n)}$ and $g_{\mu\nu}$. Although the heat kernel only resums one-loop diagrams, the effects of higher loop processes from internal gravitons and gauge particles are encoded in effective vertices, which may be verified in a Feynman diagrammatic expansion.\footnote{The same phenomenon occurs in the Euler-Heisenberg Lagrangian, cf. \cite{Schwartz:2013pla}.}

Armed with the exact effective action, we extract its logarithmic corrections in the limit where $|\vec{Q}|$ and $|\vec{q}\cdot\vec{Q}|$ are large, but $|\vec{q}|$ is small. After choosing a renormalization scheme or redefining couplings by appropriately absorbing the effective field theory cutoff, we obtain the macroscopic entropy due to the $\Phi$ field. Note that because $A_{\mu}$ and $g_{\mu\nu}$ are held fixed, their contribution to the entropy must be estimated from their separate one-loop contribution to the effective action. Additionally, one must characterize the backreaction on the gauge and gravitational fields induced by the scalar fluctuations.\footnote{We may calculate the semiclassical backreaction by solving Einstein's equations with the stress-tensor replaced by its one-loop corrected expectation value. We later show backreaction effects to be negligible for the perturbations of the renormalized effective action for the specific geometry we study.}

\section{Macroscopic Entropy}

\subsection{Contribution to Entanglement Entropy from Neutral Scalars}

We want to compute the quantum correction to the macroscopic entropy due to a gauged scalar. Let us review the calculation for a neutral, massless scalar. For each field, there are four contributions to the entropy:
\begin{equation}
    S = S_{\text{0}} + S_{\text{div}} + S_{\text{CT}} + S_{\text{fin}}
\end{equation}
\noindent where $S_{\text{0}}$ is the classical contribution to the entropy, $S_{\text{div}}$ is the UV divergent quantum correction, $S_{\text{CT}}$ is the entropy from counterterms that regulate UV divergences, and $S_{\text{fin}}$ is from finite quantum corrections to the entropy. Because the heat kernels of the individual fields add at one-loop, the total entropy is the sum of the individual fields' contributions to the entropy. Beyond one-loop, we must estimate the magnitude of entropic contributions from quantum fluctuations of the background geometry backreacting on one-another.

To compute the heat kernel of the scalar field in the \adss{} geometry, we express $\hat{H}$ as the sum of the scalar Laplacian operator on $\text{AdS}_2$ and the scalar Laplacian on $\text{S}^2$. The heat kernel factorizes as
\begin{equation}
  K(s) = K_{\text{AdS}_2}(s) K_{\text{S}^2}(s).
\end{equation}
\noindent The eigenfunctions of \stwo{} are the spherical harmonics $Y_{\ell m}(\psi,\varphi)/r_E^2$. Only the $m = 0$ eigenfunctions contribute to $K(s)$. At $\psi = 0$,
\begin{equation}
  Y_{\ell 0}(0) = \sqrt{\frac{2\ell + 1}{4\pi}},
\end{equation}
\noindent and $Y_{\ell 0}$ has eigenvalues $\ell (\ell +1)/r_E^2$. Therefore,
\begin{equation}
  K_{\stwo}(s) = \frac{1}{4\pi r_E^2} \sum\limits_{\ell = 0}^{\infty} (2\ell + 1) e^{-s\ell (\ell +1)/r_E^2}.
\end{equation}
\noindent The eigenvalues and eigenfunctions of the $\text{S}^2$ Laplacian are unaffected by the gauge covariant coupling of the $\phi$ field to the background gauge field.

The eigenfunctions of the neutral, massless scalar Laplacian on \ads{} are given in \cite{Sen11}. The full expression simplifies significantly at the origin of the \ads{} coordinate system. There, the eigenfunctions are
\begin{equation}
  f(\lambda) = \sqrt{\frac{\lambda \tanh(\lambda)}{2\pi r_E^2}},
\end{equation}
\noindent where $\lambda$ is a positive real number. The eigenvalues are
\begin{equation}
  \kappa(\lambda) = \frac{\lambda^2 + 1/4}{r_E^2}.
\end{equation}
\noindent Therefore, the heat kernel is
\begin{equation}
  K_{\ads}(s) = \frac{1}{2\pi r_E^2} \int\limits_0^{\infty} d\lambda \, \lambda \tanh(\pi\lambda) e^{-(\lambda^2 + 1/4)s/r_E^2}.
\end{equation}
\noindent We interpret $\lambda \tanh(\pi \lambda)$ as the density of states for the neutral scalar in the \ads{} background geometry.

Combining these results, we obtain the heat kernel for the neutral, massless scalar on the near-horizon background geometry 
\begin{equation}
   K(s) = \frac{1}{16\pi^2r_E^4 \overline{s}^2}\left ( 1 + \frac{\overline{s}^2}{45} \right ).
\end{equation}
\noindent Consequently, the divergent contribution to the entropy in the large $|\vec{Q}|$ limit in Planck units is
\begin{equation}\label{eq:neutralLog}
   S_{\text{div}} =  + \frac{r_E^2}{4\varepsilon^2} + \frac{1}{180}\log(\varepsilon/r_E^2).
\end{equation}
\noindent This is the \emph{exact} divergent correction to the macroscopic entropy of the black hole due to the quantum fluctuations of a neutral scalar, previously derived in \cite{Sen11}.\footnote{Up to exponentially suppressed terms and backreaction of the background fields.} The result is exact because the action is quadratic in the scalar field, and we formally solved for the heat kernel using equation ~\eqref{eq:heatFull} without a perturbative expansion.

The result matches the familiar small $s$ expansion of the heat kernel in powers and contractions of curvature invariants. The coefficients of the heat kernel expanded in $s$ are related to local quantities computed in the background geometry,
\begin{equation}
    K(s) = \sum\limits_{n=0}^{\infty} a_{2n}(R_{\mu\nu\rho\sigma}, F_{\mu\nu})s^{n-2}e^{-sm^2}, 
\end{equation}
\noindent where, for a massive scalar field in an arbitrary background geometry the coefficients are
\begin{align}
    a_0 &= \frac{1}{8\pi^2}\int d^4x\, \sqrt{\operatorname{det}\, g}\\
    a_2 &= \frac{1}{8\pi^2}\int d^4x\, \sqrt{\operatorname{det}\, g} \, \frac{1}{6} R\\
    a_4 &= \frac{1}{8\pi^2}\int d^4x\, \sqrt{\operatorname{det}\, g} \left ( 12 \nabla_{\mu}\nabla^{\mu} + 5 R^2 - 2 R_{\mu\nu}R^{\mu\nu} + 2 R_{\mu\nu\rho\sigma}R^{\mu\nu\rho\sigma} - 30 q^2 F_{\mu\nu}F^{\mu\nu} \right ).
\end{align}

For the near-horizon geometry, the constant part of $K(s)$, which is $a_4(s)$ in four-dimensions, may be reduced to
\begin{equation}
   a_4 = \frac{1}{720\pi^2} R_{\mu\nu}R^{\mu\nu} = \frac{1}{720\pi^2 r_E^2},
\end{equation}
\noindent as expected. We have set $m^2=0$ for the massless field considered in this section. For a massive field, the logarithmic divergence is damped:
\begin{equation}
    S_{\text{div,log}} = \frac{1}{180}\log\left ( \frac{\varepsilon}{m^2} \right ).
\end{equation}
\noindent If the mass is smaller the inverse radius of the extremal black hole, it is appropriate to expand the exponential for small $s$. The logarithmic divergence is a modification of the massless scalar's logarithmic divergence:
\begin{equation}\label{eq:smallmassNeutral}
    S_{\text{div,log}} = \left (  \frac{1}{180}  + \frac{1}{8} m^2 r_E^2 \right ) \log \left ( \frac{\varepsilon}{r_E^2} \right ).
\end{equation}
\noindent It may be checked \cite{Vas03} that this extra term contributes to the renormalization of the cosmological constant. When we study the gauged scalar, it is important to note that the extra divergence present in that answer takes the form of a divergent cosmological constant contribution \emph{without} any expansion of the exponential.

Because the expression for the entropy is UV divergent, we must append counterterms to the effective action to cancel the divergences. Schematically denote each counterterm by $\delta_{\mathcal{O}} \mathcal{O}$, where $\mathcal{O}$ is the operator which receives a divergent correction, and $\delta_{\mathcal{O}}$ is the counterterm. The heat kernel in the small effective mass limit has no exponential suppression. Therefore, the counterterm $\delta_{\mathcal{O}}$ introduces an arbitrary length scale $\ell$ satisfying $\varepsilon < \ell^2 < r_E^2$ to cancel the divergence in the logarithmic term that occurs when we take $\varepsilon \rightarrow 0$. Schematically, each counterterm takes the form
\begin{equation}
    \delta_{\mathcal{O}}\mathcal{O} = -\sum\limits_{n = 1}^{d/2} c^{(n)}_{\mathcal{O}}\varepsilon^{-2n} - c^{(0)}_{\mathcal{O}}\log(\ell^2/\varepsilon) = -\sum\limits_{n = 1}^{d/2} c^{(n)}_{\mathcal{O}}\varepsilon^{-2n} - c^{(0)}_{\mathcal{O}} \big[ \log(\ell_0^2/\varepsilon) + \log(\ell^2/\ell_0^2) \big].
\end{equation}
\noindent The $c^{(n)}_{\mathcal{O}}$ coefficients represent the coefficients of the divergent parts of the $\varepsilon^{-2n}$ portions of the effective action in the $\varepsilon \rightarrow 0$ limit. We introduce two arbitrary length scales $\ell$ and $\ell_0$. The length scale $\ell_0$ does not contribute to the entropy of the initial extremal black hole solution we consider, as it cancels out. However, to simplify calculations, we fix the last term in the above expression for all black hole solutions. When we renormalize both black hole solutions, this fixes both $\ell$ and $\ell_0$. Because $\ell_0$ does not appear in the entropy for the extremal black hole, choosing a renormalization condition for the initial extremal black hole fixes $\ell$. When we apply a linearized perturbation to the extremal black hole, we have chosen a convention where all terms in the entropy above change except for the last, finite counterterm. We then renormalize this black hole solution, which fixes $\ell_0$. All other black hole solutions obtained from further perturbations of the renormalized solution run with changes in the black hole parameters (charge, gauge coupling, radius) as dictated by our initially chosen renormalization conditions.

We implicitly choose a renormalization condition that exactly cancels any non-logarithmic divergences. We only discuss the logarithmically divergent counterterms in what follows, unless otherwise specifed. For the massless scalar, we must add a counterterm for the $R_{\mu\nu}R^{\mu\nu}$ operator. Its contribution to the expression for the entropy is
\begin{equation}
    S_{\text{CT,log}} = -\frac{1}{180}\log(\varepsilon/\ell^2),
\end{equation}
\noindent where $\ell$ is the arbitrary renormalization scale, in units of length. The renormalized quantum contribution to the entropy is
\begin{equation}
    S_{\text{qu}} = \frac{1}{180}\log(\ell^2/r_E^2),
\end{equation}
\noindent at extremality. If we can trust the extremal approximation near-extremality, we may simply replace the extremal radius with the outer radius of the black hole, $r_E \rightarrow r_+$. We do this when we consider small, linear perturbations to the near horizon geometry. Because $\ell$ is an ambiguous scale, we fix it by specifying our renormalization condition. For example, we may choose a condition that for a black hole of charge $\vec{Q}_0$ at extremality, the quantum corretion to the black hole entropy vanishes exactly. Because the entropy depends on the radius of the black hole, the quantum entropy of another extremal black hole of charge $\vec{Q}_0' \ne \vec{Q}_0$ or of a near-extremal black hole of charge $\vec{Q}_0$ is non-zero. In other words, the entropy runs with the radius of the black hole.

The case of a massive scalar is different. For a massive scalar with $m > 1/r_E$, we may not expand the exponential term that suppresses the heat kernel. The logarithmic contribution to the entropy is, therefore,
\begin{equation}
    S_{\text{div,log}} = \frac{1}{180}\log(\varepsilon/m^2).
\end{equation}
\noindent Up to a finite term that is independent of the black hole radius, we may choose a logarithmically divergent counterterm for $R_{\mu\nu}R^{\mu\nu}$ whose contribution to the entropy is
\begin{equation}
    S_{\text{CT,log}} = \frac{1}{180}\log(m^2/\varepsilon),
\end{equation}
\noindent which cancels the divergence exactly. There is no ambiguous renormalization scale that must be specified. This is in line with the reasoning that only massless neutral particles contribute to the entropy of large black holes. The exception is for particles with very small mass, i.e. $m < r_E^{-1}$. In that case, the renormalization to the $R_{\mu\nu}R^{\mu\nu}$ operator proceeds in the same way. An extra operator must be renormalized to absorb the extra divergent contributions to the heat kernel. The structure of the divergent terms exactly matches the contribution to the cosmological constant. We renormalize the cosmological constant to absorb its divergence \cite{Vas03}. Its counterterm contributes a logarithmically divergent term to the entropy
\begin{equation}
    S_{\text{CT,log}} = \frac{1}{360}m^4r_E^4\log(\ell^2/\varepsilon).
\end{equation}
\noindent The renormalized correction to the entropy is
\begin{equation}
    S_{\text{qu}} = \left ( \frac{1}{180} + \frac{1}{360}m^4r_E^4 \right )\log(\ell^2/r_E^2).
\end{equation}

\subsection{Entropy of Gauged Scalars}

The coupling of the gauged scalar to the background field modifies the eigenvalues and eigenfunctions of the scalar \ads{} Laplacian \cite{PioTro15}. In the near-horizon geometry, the background field strength for the $n^{\text{th}}$ gauge field in the Wick rotated spacetime is
\begin{equation}
  F^{(n)}_{\eta\theta} = i Q^{(n)} \sinh \eta.
\end{equation}
\noindent Suppose instead that the scalar is coupled to a constant background magnetic monopole field $\vec{B} = \vec{q}\cdot\vec{Q}\sin(\psi)\hat{\psi}\times\hat{\varphi}/r_E^2$. There is a continuous and discrete delta-function normalizable spectrum. The continuous eigenvalues are
\begin{equation}
  \kappa(\lambda)_B = \frac{(\lambda - \vec{q}\cdot\vec{Q})^2 + (\vec{q}\cdot\vec{Q})^2 + 1/4}{r_E^2}.
\end{equation}
\noindent The density of continuous states becomes
\begin{equation}
 \lambda \tanh(\pi\lambda) \rightarrow \lambda \frac{\sinh(2\pi\lambda)}{\cosh(2\pi\lambda) + \cos(2\pi \vec{q}\cdot\vec{Q})}.
\end{equation}
\noindent Wick rotating $\vec{q}\cdot\vec{Q} \rightarrow i \vec{q}\cdot\vec{Q}$, where $\vec{q}$ is the elementary charge vector of the $\phi$ field, we obtain the density of states for the $\phi$ field in the constant background electric field:
\begin{equation}
  \lambda \tanh(\pi \lambda) \rightarrow \lambda \frac{\sinh(2\pi\lambda)}{\cosh(2\pi\lambda) + \cosh(2\pi \vec{q} \cdot \vec{Q})}.
\end{equation}
The scalar heat kernel for the near-horizon geometry is
\begin{equation}
  K(s) = \frac{1}{8\pi^2 r_E^4} \sum\limits_{\ell = 0}^{\infty} (2\ell + 1) \int\limits_0^{\infty} d\lambda \frac{\lambda \sinh(2\pi\lambda)}{\cosh(2\pi\lambda) + \cosh(2\pi\vec{q}\cdot\vec{Q})} e^{-s(\lambda^2 + \ell(\ell+1) + \frac{1}{4} + r_E^2 m^2 - (\vec{q}\cdot\vec{Q})^2)/r_E^2}.
\end{equation}
\noindent Because the coupling $\vec{q}$ appears in the argument of a hyperbolic cosine function in the denominator of the density of states, we conclude that the resummed heat kernel represents the non-perturbative scalar field contribution to the effective action in a fixed, constant, external electric field. The result is not, however, the full quantum correction to the heat kernel. The gauge fields and gravitational field themselves contribute to the entropy. Furthermore, allowing the external gauge and gravitational fields to vary induces backreaction effects on the scalar's effective action.\footnote{The entropy due to the gauge and gravitational fields has already been tabulated in equation~\eqref{eq:SenEntropy}.}

Let us compute the divergent contributions to the effective action. Logarithmic divergences are universal and may be found in the region of integration $\varepsilon \ll s \ll r_E^2$. Therefore, we expand the resummed heat kernels for small $\overline{s} \equiv s/r_E^2$. The total heat kernel is the product of the \ads{} and \stwo{} heat kernels, weighted by a factor of $e^{-\overline{s}(r_E^2 m^2 - (\vec{q}\cdot\vec{Q})^2)}$. The expansion of the \stwo{} heat kernel is \cite{Sen11}:
\begin{equation}
  K_{\stwo}(s) = \frac{1}{4\pi r_E^2 \overline{s}} e^{\overline{s}/4} \left ( 1 + \frac{1}{12}\overline{s} + \frac{7}{480} \overline{s}^2 + \mathcal{O}(\overline{s}^3) \right ).
\end{equation}
\noindent We perform the small $\overline{s}$ expansion of the \ads{} heat kernel in its resummed form. The denominator of the \ads{} density of states has an asymptotic expansion
\begin{equation}
  \frac{1}{\cosh(2\pi\lambda) + \cosh(2\pi\vec{q}\cdot\vec{Q})} = 1 + \sum\limits_{n = 1}^{\infty} \left ( U_{n}(-\cosh(2\pi\vec{q}\cdot\vec{Q})) - U_{n-2}(-\cosh(2\pi\vec{q}\cdot\vec{Q})) \right ) e^{-2\pi n \lambda},
\end{equation}
\noindent where $U_n(x)$ is the $n^{\text{th}}$ Chebyshev polynomial of the second kind. The expansion converges when $0 \le \lambda < |\vec{q}\cdot\vec{Q}|$, regardless of the size of $|\vec{q}\cdot\vec{Q}|$. This may be checked readily by using the ratio test. The first term in the series may be evaluated directly. To evaluate the subsequent terms, we expand $e^{-\lambda^2 \overline{s}}$ for small $\overline{s}$. Denote
\begin{equation}
  \mathcal{F}_n(x) = \operatorname{Li}_n(x + \sqrt{x^2 - 1}) + \operatorname{Li}_n(x - \sqrt{x^2 - 1}),
\end{equation}
\noindent where $\operatorname{Li}_n$ is the $n^{\text{th}}$ polylogarithm. We integrate over $\lambda$ to find the \ads{} heat kernel:
\begin{align}
  K_{\ads}(s) = -\frac{e^{-\overline{s}/4}}{4\pi r_E^2 \overline{s}} \bigg [ &1 + \frac{\overline{s}}{2\pi^2} \mathcal{F}_2(-\cosh(2\pi\vec{q}\cdot\vec{Q})) \nonumber\\
  & + \overline{s}^2 \left ( \frac{7}{480} + \frac{1}{24\pi^2} \mathcal{F}_2(-\cosh(2\pi\vec{q}\cdot\vec{Q})) -  \frac{3}{4\pi^2}\mathcal{F}_4(-\cosh(2\pi\vec{q}\cdot\vec{Q})) \right ) \bigg ].
\end{align}

Through the last step, we have not made any assumptions concerning the size of $|\vec{q}\cdot\vec{Q}|$. All expansions performed have been independent of it. Now, let us take the large $|\vec{q}\cdot\vec{Q}|$ limit. We find that
\begin{align}
  \operatorname{Li}_2(-\cosh(2\pi\vec{q}\cdot\vec{Q})) &= -\frac{\pi^2}{6} - 2\pi^2 (\vec{q}\cdot\vec{Q})^2 + \mathcal{O}\left (\operatorname{sech}^2(\vec{q}\cdot\vec{Q}) \right )\\
   \operatorname{Li}_4(-\cosh(2\pi\vec{q}\cdot\vec{Q})) &= -\frac{7\pi^4}{360} - \frac{\pi^4}{3}(\vec{q}\cdot\vec{Q})^2 - \frac{2\pi^4}{3}(\vec{q}\cdot\vec{Q})^4 + \mathcal{O}\left ( \operatorname{sech}^2(\vec{q}\cdot\vec{Q}) \right ).
\end{align}
\noindent All together, the unrenormalized, large $|\vec{q}\cdot\vec{Q}|$ heat kernel is
\begin{equation}
  K(s) = \frac{e^{-\overline{s}(r_E^2 m^2 - (\vec{q}\cdot\vec{Q})^2)}}{16\pi^2 r_E^4 \overline{s}^2} \left [ 1 - \overline{s} ( \vec{q}\cdot\vec{Q} )^2 + \overline{s}^2 \left ( \frac{1}{45} + \frac{1}{6} (\vec{q}\cdot\vec{Q})^2 + \frac{1}{2}(\vec{q}\cdot\vec{Q})^4 \right ) + \mathcal{O}(\overline{s}^3) \right ].
\end{equation}
\noindent Note that our result reduces to the heat kernel of a single neutral scalar in the extremal black hole near-horizon geometry when $\vec{q} = 0$ \cite{Sen11}. Higher order terms in $\overline{s}$ contribute to finite portions of the effective action, which contribute negligibly to differences in the entropy.\footnote{One may check that the finite contributions to the entropy scale as $|\vec{q}|^{2n}|\vec{Q}|^4$ for $n >2$. However, differences between the near-extremal and extremal black hole entropies scale as $|\vec{q}|^{2n} |\vec{Q}|^2$. Because we are interested in the $|\vec{q}| \rightarrow 0$ limit, the finite contributions to the entropy are suppressed, as expected. Our logarithmic result, however, does not rely on the smallness of $|\vec{q}\cdot\vec{Q}|$, as shown explicitly in the work outlined above.} A similar, yet quantitatively different, result appears in \cite{CotShi16}.

Using the heat kernel, we may determine the logarithmic correction to the effective action, and thereby the logarithmic correction to the entropy. To connect with the weak gravity conjecture, we want to know the entropy for the resummed heat kernel, which has the $|\vec{Q}|^4$ dependence. The resummed, unrenormalized logarithmic correction to the macroscopic entropy from a single gauged scalar of mass $m$ and charge $\vec{q}$ in the large $|\vec{q}\cdot\vec{Q}|$ limit for fixed $A_{\mu}^{(n)}$, $g_{\mu\nu}$ is
\begin{equation}\label{eq:violate}
  S_{\text{div, log}} = \frac{1}{4} \left ( \frac{1}{45} + \frac{1}{6} (\vec{q}\cdot\vec{Q})^2 + \frac{1}{2} (\vec{q}\cdot\vec{Q})^4 \right ) \log(\varepsilon/(r_E^4 m^2 - r_E^2 (\vec{q}\cdot
  \vec{Q})^2)).
\end{equation}
\noindent When $r_E^2 m^2 = (\vec{q}\cdot\vec{Q})^2$, there is no exponential suppression, and the logarithmically divergent contribution to the entropy is
\begin{equation}
    S_{\text{div, log}} = \frac{1}{4} \left ( \frac{1}{45} + \frac{1}{6} ( \vec{q}\cdot\vec{Q} )^2 + \frac{1}{2} (\vec{q}\cdot\vec{Q})^4 \right ) \log ( \varepsilon/r_E^2 ).
\end{equation}

Suppose that weak gravity conjecture is satisfied but not saturated. The exponent in the resummed heat kernel before integration over $\lambda$
\begin{equation}
    \text{Exponent} = e^{-\overline{s}(\lambda^2 + r_+^2 m^2 - (\vec{q}\cdot\vec{Q})^2)}
\end{equation}
\noindent grows with increasing $\overline{s}$ for sufficiently small $\lambda$. We interpret this as an IR instability in the spectrum. The IR instability yields an imaginary contribution to the effective action \cite{PioTro15}. The magnitude of the imaginary contribution corresponds to the amount of pair production that occurs at the near-horizon geometry. We expect that one must resort to a computation of the macroscopic entropy using the Euclidean action defined on the global black hole geometry due to the instability. Additionally, we expect that it is no longer justified to work with the classical black hole background without considering how the instability backreacts on the geometry. We leave this topic for future work.

\subsection{Renormalization of Gauged Scalar Entropy}

Let us specify renormalization conditions for the initial extremal black hole solution. The black hole we consider has charge $\vec{Q}$. We assume that $|\vec{q}|$ is small, $|\vec{Q}|$ is large, and $|\vec{q}\cdot\vec{Q}|$ is large. We choose counterterms that cancel $\varepsilon^{-n}$ divergences for $n \ge 1$. The coefficient of the logarithmically divergent term is much larger than the classical contribution to the entropy. However, this does not imply that the correction for \emph{this} black hole solution is large. We choose a renormalization condition that allows us to still work in the semi-classical regime. For the perturbation we consider, we choose a renormalization condition that sets $\ell_0$ to the inverse Planck mass. Note that for large perturbations, the quantum corrections to the perturbed black hole become non-negligible.

The entanglement entropy calculated with all loop orders is given by equation~\eqref{eq:violate}. There are two important pieces of this result. First, we have the divergent term of the form 
\begin{equation}
    S_{\text{div, log}} = \frac{1}{4} \left ( \frac{1}{45} + \frac{1}{6} (\vec{q}\cdot\vec{Q})^2 + \frac{1}{2} (\vec{q}\cdot\vec{Q})^4 \right ) \log(\varepsilon/r_E^2).
\end{equation}Comparing this to equation~\eqref{eq:smallmassNeutral}, we see that this logarithmically divergent contribution resembles the contribution to the entropy from a neutral scalar field with a small mass. There are two important differences. First, the places where the small quantity $m/r_E$ appear in the expansion of the heat kernel are exactly replaced by factors of $(\vec{q}\cdot\vec{Q})^2$. This indicates that unlike the one-loop approximation to the gauged scalar heat kernel (cf appendix), there is an extra divergent contribution to the \emph{exact} heat kernel from a cosmological constant term. As with the massless neutral scalar, we may cancel the divergence from the other two terms by inserting counterterms for $R_{\mu\nu}R^{\mu\nu}$ and $F_{\mu\nu}F^{\mu\nu}$. As may be confirmed in \cite{Vas03}, the $(\vec{q}\cdot\vec{Q})^4$ requires renormalization of the cosmological constant.

The second difference is the argument of the logarithm: it depends on $(\vec{q}\cdot\vec{Q})^2$. It becomes clear what to do with the logarithmic divergence if we rewrite its contribution to the entropy in the following way:
\begin{equation}
    S_{\text{div, log}} = \frac{1}{4} \left ( \frac{1}{45} + \frac{1}{6} (\vec{q}\cdot\vec{Q})^2 + \frac{1}{2} (\vec{q}\cdot\vec{Q})^4 \right ) \left ( \log(\varepsilon/r_E^2) - \log (r_E^2 m^2) - \log\left ( 1 - \frac{(\vec{q}\cdot\vec{Q})^2}{r_E^2 m^2} \right ) \right ).
\end{equation}
\noindent Surprisingly, the divergent terms for a massive gauged scalar look like the divergent terms for a neutral scalar in the small mass limit, with the $r_E^2 m^2$ coefficient swapped for $(\vec{q}\cdot\vec{Q})^2$. The other terms are \emph{resummed, finite} corrections to the entropy. Their contributions come from an infinite sum of $(F_{\mu\nu}F^{\mu\nu})^n$-type operators. They do not require counterterms because of the lack of dependence on $\varepsilon$. Because they depend on the radius of the black hole, their contribution can only be cancelled for a \emph{specific} black hole solution. In general, they contribute a non-zero, \emph{finite} correction to the entropy at arbitrary black hole mass and charge.

We renormalize the entropy as we did for the neutral scalar in the small mass limit, with the only new feature being a $F_{\mu\nu}F^{\mu\nu}$ counterterm. The renormalized entropy is
\begin{equation}
    S_{\text{div, log}} = \frac{1}{4} \left ( \frac{1}{45} + \frac{1}{6} (\vec{q}\cdot\vec{Q})^2 + \frac{1}{2} (\vec{q}\cdot\vec{Q})^4 \right ) \left ( \log(\ell^2/r_E^2) - \log (r_E^2 m^2) - \log\left ( 1 - \frac{(\vec{q}\cdot\vec{Q})^2}{r_E^2 m^2} \right ) \right ).
\end{equation}
\noindent We specify a renormalization condition that sets the finite contributions to the entropy from resummation as well as the divergent terms equal to zero for a black hole of fixed charge $\vec{Q}$ exactly at extremality. This removes the ambiguity for the renormalization scale $\ell$ and removes all divergences.

\section{Violating the Second Law}

\subsection{Setup}

The second law states that entropy increases under any physical process:
\begin{equation}
  dS \ge 0.
\end{equation}
\noindent For healthy semi-classically treated effective field theories in curved space, the second law has been proven within various settings, e.g. \cite{Wal11}. The entropy $S$ has contributions from the classical and quantum parts of the effective action: the Bekenstein-Hawking area term as well as quantum corrections from the macroscopic fields:
\begin{equation}
   \mathcal{S} = - W_{\text{eff}} =  -(S_0 + S_{\text{quant}}),
\end{equation}
\begin{equation}
    S_{\text{quant}} = S_{\text{div}} + S_{\text{CT}},
\end{equation}
\noindent where we have neglected subleading finite corrections in $S_{\text{quant}}$. In our physical scenario, the entropy changes when the black hole consumes a neutral particle because the black hole's radius increases.  Let subscript $f$ denote final quantities, subscript $i$ initial quantities, $A \equiv 4 \pi r_+^2$ the area, and $S_{\text{quant}}$ the quantum entropy correction. Then
\begin{equation}\label{eq:bound}
  S_{0,f} - S_{0,i} \ge S_{\text{quant},i} - S_{\text{quant},f}
\end{equation}
\noindent follows from the second law.

In our thought experiment, we let a single neutral particle crosses the black hole horizon with energy $E$. This induces a linearized perturbation of the extremal black hole geometry. The black hole charge remains fixed. The initial black hole entropy has been set to its classical entropy $S_0$ and energy $E_0$ values by choosing the appropriate renormalization condition:
\begin{equation}
 \Delta W_{\text{eff},i} = 0,
\end{equation}
\begin{equation}
 S_{0,i} = \pi r_E^2 = \pi |\vec{Q}|^2,
\end{equation}
\begin{equation}
 E_{0,i} = r_E = M = |\vec{Q}|.
\end{equation}
\noindent By conservation of energy,  the black hole mass shifts to
\begin{equation}
  M_f = M + \delta M.
\end{equation}
\noindent Then
\begin{equation}
  M_f > |\vec{Q}|.
\end{equation}

The perturbed black hole receives a quantum contribution to its entropy because we have already specified fixed counterterms for the effective action of the black hole and the divergent contributions to the entropy depend on the radius of the black hole. The quantum contribution to the entropy may be mathematically traced to the fact that it runs with the radius of the horizon of the black hole. Because the quantum contribution to the \emph{exact} contribution of the gauged scalar to the black hole entropy modulo backreaction outscales the classical contribution, we expect large perturbations to the classical geometry may induce large quantum backreaction. We therefore consider small perturbations to the geometry and write the near-extremal radius $r_+$ of the perturbed black hole as
\begin{equation}
	r_+^2 = r_E^2 + \delta r^2.
\end{equation}
\noindent Note that in what follows we only consider the gauged scalar matter sector and small perturbations to the geometry in our second law analysis. We justify our result in the next section by demonstrating that effects from all other fields are subleading at one-loop and suppressed at higher loop orders and that quantum backreaction may be neglected for small perturbations of the geometry.

The only modification to the entropy at the level of linearized backreaction arises from the change in the near-horizon electric field, which shifts from $2|\vec{q}|^2\cos^2(\vartheta) \equiv (\vec{q}\cdot\vec{Q})^2/r_E^2$ to $(\vec{q}\cdot\vec{Q})^2/r_+^2$. The logarithmic correction to the classical entropy of the new black hole is\footnote{In Planck units $\mpl = 1$.}
\begin{equation}
  S_{\text{quant}} = \frac{1}{4} \left [ \frac{1}{45} + \frac{1}{6} (\vec{q}\cdot\vec{Q})^2 + \frac{1}{2} (\vec{q}\cdot\vec{Q})^4 + \mathcal{O}\left (  \operatorname{sech}^2[(\vec{q}\cdot\vec{Q})^2] \right ) \right ] \log\left ( \frac{m^2 r_+^4 - (\vec{q}\cdot\vec{Q})^2 r_+^2}{m^2 r_E^4 - (\vec{q}\cdot\vec{Q})^2 r_E^2} \right ).
\end{equation}
\noindent Ignoring exponentially suppressed contributions and keeping only the $\mathcal{O}(|\vec{Q}|^2)$ or higher terms, the bound equation~\eqref{eq:bound} becomes
\begin{equation}
   | \vec{Q}  |^2 \le \frac{32\pi}{|\vec{q}|^4 \cos^4(\vartheta)}\frac{m^2 - |\vec{q}|^2 \cos^2(\vartheta)}{2 m^2 - |\vec{q}|^2 \cos^2(\vartheta)} - \frac{1}{3}\frac{1}{|\vec{q}|^2 \cos^2(\vartheta)}.,
\end{equation}
\noindent where $\vartheta$ is the angle between $\vec{Q}$ and $\vec{q}$. The bound applies to particles violating or saturating ($m^2 = |\vec{q}|^2\cos^2(\vartheta)$) the conjecture. The dependence on $\delta r^2$ cancels on both sides of the bound for small $\delta r^2$. We may always choose an initially large, extremal black hole such that we violate the bound. A conservative interpretation of the result is that there is a maximum charge allowed in the macroscopic theories considered. This would require the appearance of some instability for large black holes. There is no evidence that this is the case, however, as we discuss in the next section.


It is natural to wonder if our result is nullified when instanton tunnelling, quantum backreaction, and effects from other fields are accounted for. The answer is no. Because the differences in quantum contributions to the entropy dominate differences in classical contributions to the entropy, large black holes are stable against splitting into multiple black holes whose charge adds up to the charge of the large black hole. This is more general than the statement that no Schwinger pair production occurs for extremal black holes formed in theories violating the weak gravity conjecture. Quantum effects dominate differences in entropy, but do not dominate the classical expressions for the entropy themselves for a suitable renormalization condition. One may worry that changes in energy, related to the backreaction of the quantum fields on the classically perturbed geometry, are important too. In fact, quantum backreaction on the black hole mass only appears at $\mathcal{O}((\delta r^2)^2)$. Moreover, as aforementioned and cited, contributions at one-loop from other massless fields, such as massless matter, other gauge fields, and the gravitational field, are always subleading with respect to the classical entropy of the black hole. For large black holes, only the one-loop answer contributes in the large $|\vec{Q}|$ limit: higher loop contributions are suppressed by factors of inverse radii of the black hole. We leave these results to the appendix.

We cannot emphasize enough that \emph{the answer we have obtained is an exact answer that extends beyond the one-loop approximation}: higher loop factors have in effect been resummed because we computed the full partition function for the scalar field. The reason we could do this is because the action is quadratic in the scalar field, so the Euclidean path integral reduces to a Gaussian integral. Because of the special geometry of the near horizon region, we were able to compute this result analytically. Any error in our result is of order $\mathcal{O}(\text{sech}(|\vec{q}\cdot\vec{Q}|))$, which is suppressed in the large $|\vec{q}\cdot\vec{Q}|$ limit. As shown in the appendix, quantum backreaction does not affect the classical geometry at order $\delta r^2$ after the neutral particle crosses the black hole horizon. This is the only effect that is not explicitly captured by our exact computation.

\section{Consistency Checks}

\subsection{Subleading Contributions from Neutral Matter, Gauge Fields, and Gravitational Field at All Loop Orders}

Our expression for the exact heat kernel of the scalar field indicates a second law violation. We have not included effects from the two other fields present: the $U(1)^N$ gauge field and the gravitational field. This is because these contributions are subleading. The reason that the scalar had such a large contribution to its entropy is because it couples to the background gauge field. Therefore, the action for the $\phi$ field includes contributions from positive powers of the background black hole charge. This is not the case for the gravitational and gauge actions.

Let us first consider the one-loop contributions to the heat kernel from the gauge and gravitational fields. The total heat kernel for the full theory is the sum of the individual heat kernels, so we can consider each field separately. At one-loop, we only need to consider the quadratic action for each field. The one-loop expression for $N$ $U(1)$ vector fields and the gravitational field has already been known for some time, calculated by Sen in \cite{Sen11}:
\begin{equation}
    S = \frac{A}{4} - \frac{1}{180} \left ( 964 + 62 N \right ) \log(A).
\end{equation}
\noindent Note that the quantum correction is \emph{subleading}. Moreover, Sen \emph{et al.} demonstrate in \cite{Sen11} that higher loop contributions are suppressed in the large black hole mass limit. Therefore, the one-loop result for the macroscopic answer suffices.

\subsection{Suppression of Quantum Backreaction for Small Classical Perturbations}

In our analysis, we choose renormalization conditions such that the quantum correction to the extremal black hole entropy with charge $\vec{Q}$ is absorbed into the tree-level, classical value for the entropy. The linear perturbation to the black hole induced by a neutral particle crossing the horizon causes the quantum entropy to run, because the entropy depends on the radius of the black hole. The quantum correction to the entropy of the perturbed black hole solution is smaller than the classical entropy of the perturbed black hole. However, the difference between the classical entropies of the initial and final black holes is smaller than the difference in quantum corretions to the entropy. It is for this reason that the second law is violated. We use the exact expression for the scalar field effective action for fixed, external classical backgrond fields, accounting only for classical backreaction. Here we provide a back of the envelope argument that quantum backreaction does not modify our result at $\mathcal{O}(\delta r^2)$.

The mass of the black hole may be expressed via the first law as
\begin{equation}
    M = T S + |\vec{Q}|,
\end{equation}
\noindent where $T$ is the temperature, $S$ is the entropy, $\vec{Q}$ is the charge, and we have set the chemical potential to one. We assume that quantum corrections to all quantities written above are factored into this formula. For a stationary, charged black hole, these are the only sources that can contribute to the black hole mass.

Let us consider backreaction on the charge of the black hole. The charge of the perturbed black hole receives quantum corrections that are of order
\begin{equation}
    \vec{Q}_{\text{qu}} \propto -|\vec{q}|^4 |\vec{Q}|^0 \delta r^2.
\end{equation}
\noindent In the small $|\vec{q}|$ limit, we may assume that these are subleading and ignore these corrections only if the quantum corrections to the mass do not dominate.

The perturbed black hole has a classical correction to its mass proportional to the temperature of the black hole. The classical temperature is
\begin{equation}
    T = \frac{1}{2\pi}\left ( \frac{1}{r_+} - \frac{|\vec{Q}|^2}{r_+^3} \right ),
\end{equation}
\noindent which evalutes to
\begin{equation}
    T = \frac{\delta r^2}{|\vec{Q}|^3} + \mathcal{O}((\delta r^2)^2).
\end{equation}
\noindent It may be checked that quantum corrections do not modify this order of magnitude estimate. The thermal contribution to the mass of the black hole has a classical and quantum component. The classical contribution arises from the classical entropy:
\begin{equation}
    M_{cl} = 4\pi T r_+^2 = 4\pi \frac{\delta r^2}{|\vec{Q}|} + \mathcal{O}((\delta r^2)^2).
\end{equation}
\noindent The quantum correction to the entropy is proportional to $\mathcal{O}(\delta r^2)|\vec{q}|^4|\vec{Q}|^4$. Therefore, the quantum correction to $M$ is proportional to $(\delta r^2)^2$:
\begin{equation}
    M_{qu} \propto - |\vec{q}|^4|\vec{Q}| \mathcal{O}((\delta r^2)^2).
\end{equation}
\noindent In the small $\delta r^2$ expansion, this is smaller than the classical backreaction near-extremality. We conclude that we may ignore quantum backreaction effects in our thought experiment. A full analysis should utilize the semi-classical Einstein equations. We leave this to future work.

\subsection{Stability of the Near Horizon Geometry}

A black hole with charge $\vec{Q}$ is not the only classical geometry asymptotic to the \adss{} in the near-horizon limit. Other geometries that contribute to the path integral are multi-black hole solutions, where the total charge of the black holes equals $\vec{Q}$. When the weak gravity conjecture is satisfied, tunneling processes may occur in which the initial \adss{} near-horizon geometry fragments into multiple \adss{} geometries. The simplest example is the Brill instanton, wherein one initial \adss{} space tunnels into two disconnected spaces. Let us review the calculation using the classical piece of the effective action first, following \cite{MalMic98}. For simplicity of presentation, we work with a $U(1)$ gauge group in the remainder of this section. The background gauge field in the two black hole solution is
\begin{equation}
     A_{t}(\vec{x}) = \frac{Q_1}{|\vec{x} - \vec{x}_1|} + \frac{Q_2}{|\vec{x} - \vec{x}_2|}.
\end{equation}
\noindent Further details may be found in \cite{MalMic98}. The instanton action is half the negative difference of the initial and final black hole entropies,\footnote{The factor of $\frac{1}{2}$ appears because the transition probability between solutions is proportional to $e^{-\Delta S}$.}
\begin{equation}
    S_{\text{inst}} = -\frac{1}{2}\Delta S.
\end{equation}

Consider the Bekenstein-Hawking term without quantum corrections. The Brill instanton action is
\begin{equation}
    S_{\text{inst}} = \pi Q_1 Q_2.
\end{equation}
\noindent The transition amplitude from the charge $Q$ black hole to the $Q_1$ and $Q_2$ charged black holes is
\begin{equation}
    A_{Q \rightarrow Q_1 + Q_2} \propto e^{\frac{1}{2}\Delta S},
\end{equation}
\noindent up to normalization. Consequently, the transition probability is
\begin{equation}
    P_{Q \rightarrow Q_1 + Q_2} \propto e^{-{\pi}Q_1 Q_2}.
\end{equation}
\noindent The probability is less than one for non-zero $Q_1$ and $Q_2$, as expected.\footnote{It is implicit in what follows that $Q_1$ and $Q_2$ have the same sign.} When $Q_{1} \rightarrow q$, this answer represents the probability amplitude for brane-antibrane production, i.e. Schwinger pair production. Now consider quantum corrections to the macroscopic entropy from matter neutral under the $U(1)$ gauge symmetry. The logarithmic terms are subleading. Therefore, the instanton action is still positive, because $S(Q) \ge S(Q_1) + S(Q_2)$. We interpret this to mean that large black holes dominate the Euclidean path integral, with an exponentially suppressed probability that the black holes fragment into multi-black hole solutions. Note that fragmentation and Schwinger pair production would preclude us from maintaining a sufficient level of control over the process we consider.

Now consider the quantum corrected black hole entropy when the weak gravity conjecture is violated or saturated for a non-supersymmetric gauged scalar. In the large charge limit, one may verify that
\begin{equation}
    S(Q) < S(Q_1) + S(Q_2).
\end{equation}
\noindent Consequently, the physical instanton process is not fragmentation; rather, it is black hole growth from an initial two black hole state to a single black hole final state. This is consistent with the $Q_{1} \rightarrow q$ limit: there is no pair production. Similarly, there is no black hole fragmentation. Instead, the correct instanton action corresponds to two initial \adss{} states transitioning into one final \adss{} states. In the $Q_{1} \rightarrow q$ limit, this is a process akin to the thought experiment in the previous section. The combined state of the black hole and a particle eventually transitions to a final state where the black hole consumes the charged particle. In conclusion, the large-charged black hole in our setup does not fragment into a multitude of black holes or charged particles. This is in accord with the kinematics arguments presented in the introduction. The theory contains only subextremal objects in its spectrum, so the extremal black hole has no decay channels. Moreover, the instanton analysis implies that the Euclidean path integral is dominated by small black hole classical saddle points, i.e. remnants. We claim that the absence of a decay channel affords us sufficient control over the process we consider.

It is clear now that black hole growth is the favored physical process in theories violating the weak gravity conjecture, at least in situations where backreaction can be neglected. We speculate that the reversal of the Brill instanton violates unitarity. Renormalize the large extremal black hole effective action. One may tune the effective action such that the entropy is positive, despite the seemingly large quantum correction. However, no mechanism exists within the IR theory that prevents the black hole from continuing to grow unbounded. When the black hole grows, the decreasing quantum correction outcompetes the increasing Bekenstein-Hawking term. Counterintuitively, larger black holes hide fewer microscopic states behind the horizon than smaller black holes. Because growth may occur without bound, the entropy eventually becomes negative, indicating that the black hole contains less than one microscopic state. We expect that this behavior is forbidden in a unitary theory. Therefore, we speculate that the scalar violating the weak gravity conjecture is secretly non-unitarity, even at the level of effective field theory. 

\section{Conclusion and Future Work}

This paper comprises a proof of the weak gravity conjecture, obtained from studying the macroscopic entropy of gauged scalars on a semiclassical near-extremal black hole background. Our choice of renormalization conditions allows us to safely neglect non-linear metric backreaction. The quantum corrected entropy violates the second law if the conjecture is not satisfied. When the conjecture is satisfied, the black hole near extremality decays rapidly due to Schwinger pair production, which allows the theory to evade the troubling thermodynamic violation. Therefore, we establish that it is necessary that a weak gravity conjecture is obeyed.\footnote{We leave it to future work to determine if it is \emph{sufficient}.}. Our calculation demonstrates that entropy inequalities may discriminate between effective field theories that live in the landscape versus the swampland. Although effective field theories that violate the weak gravity conjecture do not obviously violate unitarity, positivity, or causality, the violation of the second law indicates that \emph{some} sickness lurks within them. In conclusion, we propose that a violation of the second law modulo backreaction indicates an IR obstruction to a UV completion in a unified theory \cite{Adams:2006sv}.

Our analysis does not truly address weak gravity in effective field theory or on arbitrary perturbations of the black hole background. We only consider the minimally gauged, minimally coupled quadratic action of the $D = 4$ gauged scalar. A follow-up paper \cite{FisMog17b} bridges the gap: we address the conjecture in arbitrary dimensions and non-minimal interactions, including non-renormalizable terms. We limited our analysis here to the minimal quadratic action for ease of presentation and because we could obtain an exact result. We extend our result to actions with higher dimension operators and to actions with multiple scalars in \cite{FisMog17b}. In particular, we prove the generalized electric weak gravity conjecture of \cite{CheRem14} in our follow-up paper.

It would be worthwhile to extend our methodology to arbitrary $p$-form gauge fields. For example, while it is expected that there is a weak gravity conjecture for $p > 1$, it is unclear if $p = 0$ axions are subject to a weak gravity conjecture. If they are, then there are direct implications for inflationary model building. In particular, large field axion inflation would violate the $p = 0$ weak gravity conjecture \cite{Heidenreich:2015wga}.

Although our results directly apply to the weak gravity conjecture, they might also apply to the Ooguri-Vafa conjecture \cite{OogVaf16}.\footnote{See also \cite{FreKle16}.} Ooguri and Vafa claim that there are no stable non-supersymmetric AdS vacuua whose cosmological constant is supported by a flux. If true, then the conjecture has serious implications for non-supersymmetric AdS/CFT. Large-$N$ brane constructions and Kaluza-Klein compactifications include extremal particles in the bulk spectrum. Our result demonstrates a conflict between thermodynamics and non-supersymmetric, gauged extremal particles, suggesting a route to proving the conjecture.

The extensions aforementioned do not capture the full potential of our methodology. We propose that the armamentarium of entropy technology at our disposal may define new, undiscovered constraints on effective field theories compatible with quantum gravity. Our follow-up paper provides minor evidence in favor of the proposal \cite{FisMog17b}. The power of the methodology lies within the relative ease of calculating macroscopic entropy of IR field content in semi-classical gravitational backgrounds. One may remain agnostic as to the full UV completion of the effective theory. Nonetheless, if the effective theory violates known entropy inequalities in the IR, then there exists some obstruction to a UV completion.

\section*{Acknowledgements}

The authors are especially grateful to Matt Reece for thoroughly reviewing a late-stage draft of the manuscript and Aron Wall for extensive, thoughtful discussion and debate. We also thank Raphael Bousso, Venkatesa Chandrasekar, Netta Engelhardt, Illan Halpern, Petr Ho\v{r}ava, Juan Maldacena, Arvin Moghaddam, Fabio Sanches, and Ziqi Yan for conversations during various stages of preparation of this manuscript. The authors are also very grateful to John Brown and Xiaobei Wei for their hospitality at our second home, Sophie's Cuppa Tea.

C.M. is supported by a National Science Foundation Graduate Research Fellowship. The work of Z.F. is supported in part by the Berkeley Center for Theoretical Physics, by the National Science Foundation (award numbers 1214644, 1316783, and 1521446), by fqxi grant RFP3-1323, and by the US Department of Energy under Contract DE-AC02-05CH11231. This work was completed at the Aspen Center for Physics, which is supported by National Science Foundation grant PHY-1066293.

\begin{appendix}

\section{One-Loop Calculation}

The exact heat kernel for the minimally gauged scalar in the presence of fixed external background fields does not match the expected one-loop result. In the one-loop heat kernel Suppose that $m \gg q \mpl$. Then we may expand the $s(\vec{q}\cdot\vec{Q})^2$ part of the argument of the exponent:
\begin{equation}
   K(s) \approx \frac{1}{16\pi^2 r_E^4 \overline{s}^2}\left [ 1 + \overline{s}^2 \left ( \frac{1}{45} + \frac{1}{6} (\vec{q}\cdot\vec{Q})^2 \right ) + \mathcal{O}(\overline{s}^4) \right ] e^{-\overline{s}r_E^2 m^2}.
\end{equation}
\noindent This is exactly what one would obtain in the geometric expansion of the heat kernel in the large $|\vec{Q}|$ limit:
\begin{align}
    K(s) &\approx \frac{1}{16\pi^2 \overline{s}^2} \left( 1 + \frac{\overline{s}}{6} R 
     + \frac{\overline{s}^2}{45} R_{\mu\nu}R^{\mu\nu} + \frac{1}{6} q^2 F_{\mu\nu}F^{\mu\nu} \right) e^{-\overline{s}r_E^2 m^2}.
\end{align}
\noindent The $(\vec{q}\cdot\vec{Q})^4$ term is cancelled by the background gauge field term when we expand the exponential. Therefore, in the large mass limit, one may verify that only a $q^2 F_{\mu\nu}F^{\mu\nu}$ counterterm is required to cancel the divergence due to powers of $\vec{q}\cdot\vec{Q}$ that appear in the final result, which can be seen by performing the small $s$ expansion or, likewise, expanding the exponent in our exact result.\footnote{It can be verified that it is only in this limit that the approximation of the integrand made in \cite{CotShi16} is justified. Because this is not a focus of our paper, we refrain from providing further commentary in this paper on this detail.}

\end{appendix}

\bibliographystyle{utcaps}

\begin{thebibliography}{10}

\bibitem{ArkMot06}
N.~Arkani-Hamed, L.~Motl, A.~Nicolis, and C.~Vafa, ``The string landscape,
  black holes and gravity as the weakest force,'' {\em JHEP} {\bfseries 06}
  (2007) 060, \href{http://arxiv.org/abs/hep-th/0601001}{{\ttfamily
  hep-th/0601001}}.

\bibitem{Heidenreich:2016aqi}
B.~Heidenreich, M.~Reece, and T.~Rudelius, ``{Evidence for a Lattice Weak
  Gravity Conjecture},''
\href{http://arxiv.org/abs/1606.08437}{{\ttfamily arXiv:1606.08437 [hep-th]}}.

\bibitem{Banks:2006mm}
T.~Banks, M.~Johnson, and A.~Shomer, ``{A Note on Gauge Theories Coupled to
  Gravity},'' \href{http://dx.doi.org/10.1088/1126-6708/2006/09/049}{{\em JHEP}
  {\bfseries 09} (2006) 049},
\href{http://arxiv.org/abs/hep-th/0606277}{{\ttfamily arXiv:hep-th/0606277
  [hep-th]}}.

\bibitem{Cas08}
H.~Casini, ``Relative entropy and the Bekenstein bound,''
  \href{http://dx.doi.org/10.1088/0264-9381/25/20/205021}{{\em
  Class.Quant.Grav.} {\bfseries 25} (2008) 205021},
\href{http://arxiv.org/abs/0804.2182}{{\ttfamily arXiv:0804.2182 [hep-th]}}.

\bibitem{CheRem14}
C.~Cheung and G.~N. Remmen, ``{Naturalness and the Weak Gravity Conjecture},''
  \href{http://dx.doi.org/10.1103/PhysRevLett.113.051601}{{\em Phys. Rev.
  Lett.} {\bfseries 113} (2014) 051601},
\href{http://arxiv.org/abs/1402.2287}{{\ttfamily arXiv:1402.2287 [hep-ph]}}.

\bibitem{CheRem14b}
C.~Cheung and G.~N. Remmen, ``{Infrared Consistency and the Weak Gravity
  Conjecture},'' \href{http://dx.doi.org/10.1007/JHEP12(2014)087}{{\em JHEP}
  {\bfseries 12} (2014) 087},
\href{http://arxiv.org/abs/1407.7865}{{\ttfamily arXiv:1407.7865 [hep-th]}}.

\bibitem{Har15}
D.~Harlow, ``{Wormholes, Emergent Gauge Fields, and the Weak Gravity
  Conjecture},'' \href{http://dx.doi.org/10.1007/JHEP01(2016)122}{{\em JHEP}
  {\bfseries 01} (2016) 122},
\href{http://arxiv.org/abs/1510.07911}{{\ttfamily arXiv:1510.07911 [hep-th]}}.

\bibitem{Heidenreich:2015wga}
B.~Heidenreich, M.~Reece, and T.~Rudelius, ``{Weak Gravity Strongly Constrains
  Large-Field Axion Inflation},''
  \href{http://dx.doi.org/10.1007/JHEP12(2015)108}{{\em JHEP} {\bfseries 12}
  (2015) 108},
\href{http://arxiv.org/abs/1506.03447}{{\ttfamily arXiv:1506.03447 [hep-th]}}.

\bibitem{Heidenreich:2015nta}
B.~Heidenreich, M.~Reece, and T.~Rudelius, ``{Sharpening the Weak Gravity
  Conjecture with Dimensional Reduction},''
  \href{http://dx.doi.org/10.1007/JHEP02(2016)140}{{\em JHEP} {\bfseries 02}
  (2016) 140},
\href{http://arxiv.org/abs/1509.06374}{{\ttfamily arXiv:1509.06374 [hep-th]}}.

\bibitem{BouCas14a}
R.~Bousso, H.~Casini, Z.~Fisher, and J.~Maldacena, ``Proof of a Quantum Bousso
  Bound,'' \href{http://dx.doi.org/10.1103/PhysRevD.90.044002}{{\em Phys.Rev.}
  {\bfseries D90} no.~4, (2014) 044002},
\href{http://arxiv.org/abs/1404.5635}{{\ttfamily arXiv:1404.5635 [hep-th]}}.

\bibitem{BouCas14b}
R.~Bousso, H.~Casini, Z.~Fisher, and J.~Maldacena, ``Entropy on a null surface
  for interacting quantum field theories and the Bousso bound,''
  \href{http://dx.doi.org/10.1103/PhysRevD.91.084030}{{\em Phys.Rev.}
  {\bfseries D91} no.~8, (2015) 084030},
\href{http://arxiv.org/abs/1406.4545}{{\ttfamily arXiv:1406.4545 [hep-th]}}.

\bibitem{BouFis15a}
R.~Bousso, Z.~Fisher, S.~Leichenauer, and A.~C. Wall, ``A Quantum Focussing
  Conjecture,'' \href{http://arxiv.org/abs/1506.02669}{{\ttfamily
  arXiv:1506.02669 [hep-th]}}.

\bibitem{BouFis15b}
R.~Bousso, Z.~Fisher, J.~Koeller, S.~Leichenauer, and A.~C. Wall, ``{Proof of
  the Quantum Null Energy Condition},''
  \href{http://dx.doi.org/10.1103/PhysRevD.93.024017}{{\em Phys. Rev.}
  {\bfseries D93} no.~2, (2016) 024017},
\href{http://arxiv.org/abs/1509.02542}{{\ttfamily arXiv:1509.02542 [hep-th]}}.

\bibitem{Banerjee:2011jp}
S.~Banerjee, R.~K. Gupta, I.~Mandal, and A.~Sen, ``{Logarithmic Corrections to
  N=4 and N=8 Black Hole Entropy: A One Loop Test of Quantum Gravity},''
  \href{http://dx.doi.org/10.1007/JHEP11(2011)143}{{\em JHEP} {\bfseries 11}
  (2011) 143},
\href{http://arxiv.org/abs/1106.0080}{{\ttfamily arXiv:1106.0080 [hep-th]}}.

\bibitem{Bhattacharyya:2012wz}
S.~Bhattacharyya, B.~Panda, and A.~Sen, ``{Heat Kernel Expansion and Extremal
  Kerr-Newmann Black Hole Entropy in Einstein-Maxwell Theory},''
  \href{http://dx.doi.org/10.1007/JHEP08(2012)084}{{\em JHEP} {\bfseries 08}
  (2012) 084},
\href{http://arxiv.org/abs/1204.4061}{{\ttfamily arXiv:1204.4061 [hep-th]}}.

\bibitem{Sen:2011ba}
A.~Sen, ``{Logarithmic Corrections to N=2 Black Hole Entropy: An Infrared
  Window into the Microstates},''
  \href{http://dx.doi.org/10.1007/s10714-012-1336-5}{{\em Gen. Rel. Grav.}
  {\bfseries 44} no.~5, (2012) 1207--1266},
\href{http://arxiv.org/abs/1108.3842}{{\ttfamily arXiv:1108.3842 [hep-th]}}.

\bibitem{Sen11}
S.~Banerjee, R.~K. Gupta, and A.~Sen, ``{Logarithmic Corrections to Extremal
  Black Hole Entropy from Quantum Entropy Function},''
  \href{http://dx.doi.org/10.1007/JHEP03(2011)147}{{\em JHEP} {\bfseries 03}
  (2011) 147},
\href{http://arxiv.org/abs/1005.3044}{{\ttfamily arXiv:1005.3044 [hep-th]}}.

\bibitem{Sen:2008vm}
A.~Sen, ``{Quantum Entropy Function from AdS(2)/CFT(1) Correspondence},''
  \href{http://dx.doi.org/10.1142/S0217751X09045893}{{\em Int. J. Mod. Phys.}
  {\bfseries A24} (2009) 4225--4244},
\href{http://arxiv.org/abs/0809.3304}{{\ttfamily arXiv:0809.3304 [hep-th]}}.

\bibitem{Sen:2012dw}
A.~Sen, ``{Logarithmic Corrections to Schwarzschild and Other Non-extremal
  Black Hole Entropy in Different Dimensions},''
  \href{http://dx.doi.org/10.1007/JHEP04(2013)156}{{\em JHEP} {\bfseries 04}
  (2013) 156},
\href{http://arxiv.org/abs/1205.0971}{{\ttfamily arXiv:1205.0971 [hep-th]}}.

\bibitem{CotShi16}
W.~Cottrell, G.~Shiu, and P.~Soler, ``{Weak Gravity Conjecture and Extremal
  Black Holes},''
\href{http://arxiv.org/abs/1611.06270}{{\ttfamily arXiv:1611.06270 [hep-th]}}.

\bibitem{PioTro15}
B.~Pioline and J.~Troost, ``{Schwinger pair production in AdS(2)},''
  \href{http://dx.doi.org/10.1088/1126-6708/2005/03/043}{{\em JHEP} {\bfseries
  03} (2005) 043},
\href{http://arxiv.org/abs/hep-th/0501169}{{\ttfamily arXiv:hep-th/0501169
  [hep-th]}}.

\bibitem{Hod17}
S.~Hod, ``{A proof of the weak gravity conjecture},''
\href{http://arxiv.org/abs/1705.06287}{{\ttfamily arXiv:1705.06287 [gr-qc]}}.

\bibitem{Schwartz:2013pla}
M.~D. Schwartz, {\em {Quantum Field Theory and the Standard Model}}.
\newblock Cambridge University Press,
2014.
\newblock

\bibitem{Vas03}
D.~V. Vassilevich, ``Heat kernel expansion: User's manual,''
  \href{http://dx.doi.org/10.1016/j.physrep.2003.09.002}{{\em Phys.Rept.}
  {\bfseries 388} (2003) 279--360},
\href{http://arxiv.org/abs/hep-th/0306138}{{\ttfamily arXiv:hep-th/0306138
  [hep-th]}}.

\bibitem{Solodukhin11}
S.~N. Solodukhin, ``Entanglement entropy of black holes,'' {\em Living
  Rev.Rel.} {\bfseries 14} (2011) 8,
\href{http://arxiv.org/abs/1104.3712}{{\ttfamily arXiv:1104.3712 [hep-th]}}.

\bibitem{Zaslavsky:1997ha}
O.~B. Zaslavsky, ``{Geometry of nonextreme black holes near the extreme
  state},'' \href{http://dx.doi.org/10.1103/PhysRevD.56.2188,
  10.1103/PhysRevD.59.069901}{{\em Phys. Rev.} {\bfseries D56} (1997)
  2188--2191}, \href{http://arxiv.org/abs/gr-qc/9707015}{{\ttfamily
  arXiv:gr-qc/9707015 [gr-qc]}}.
[Erratum: Phys. Rev.D59,069901(1999)].

\bibitem{Pathak:2016vfc}
A.~Pathak, A.~P. Porfyriadis, A.~Strominger, and O.~Varela, ``{Logarithmic
  corrections to black hole entropy from Kerr/CFT},''
  \href{http://dx.doi.org/10.1007/JHEP04(2017)090}{{\em JHEP} {\bfseries 04}
  (2017) 090},
\href{http://arxiv.org/abs/1612.04833}{{\ttfamily arXiv:1612.04833 [hep-th]}}.

\bibitem{Wal11}
A.~C. Wall, ``A proof of the generalized second law for rapidly changing fields
  and arbitrary horizon slices,''
  \href{http://dx.doi.org/10.1103/PhysRevD.87.069904,
  10.1103/PhysRevD.85.104049}{{\em Phys.Rev.} {\bfseries D85} no.~6, (2012)
  104049},
\href{http://arxiv.org/abs/1105.3445}{{\ttfamily arXiv:1105.3445 [gr-qc]}}.

\bibitem{FisMog17b}
Z.~Fisher and C.~J. Mogni (to appear).

\bibitem{CriSan} 
  T.~Crisford and J.~E.~Santos,
  ``Violating the Weak Cosmic Censorship Conjecture in Four-Dimensional Anti–de Sitter Space,''
  \href{https://journals.aps.org/prl/abstract/10.1103/PhysRevLett.118.181101}{\em{Phys.Rev.Lett.}\  {\bf 118}, no. 18, 181101 (2017)}, 
  \href{https://arxiv.org/abs/1702.05490}{\ttfamily arXiv:1702.05490 [hep-th]}.

\bibitem{Adams:2006sv}
A.~Adams, N.~Arkani-Hamed, S.~Dubovsky, A.~Nicolis, and R.~Rattazzi,
  ``{Causality, analyticity and an IR obstruction to UV completion},''
  \href{http://dx.doi.org/10.1088/1126-6708/2006/10/014}{{\em JHEP} {\bfseries
  10} (2006) 014},
\href{http://arxiv.org/abs/hep-th/0602178}{{\ttfamily arXiv:hep-th/0602178
  [hep-th]}}.

\bibitem{OogVaf16}
H.~Ooguri and C.~Vafa, ``{Non-supersymmetric AdS and the Swampland},''
\href{http://arxiv.org/abs/1610.01533}{{\ttfamily arXiv:1610.01533 [hep-th]}}.

\bibitem{FreKle16}
B.~Freivogel and M.~Kleban, ``{Vacua Morghulis},''
\href{http://arxiv.org/abs/1610.04564}{{\ttfamily arXiv:1610.04564 [hep-th]}}.

\bibitem{MalMic98}
J.~M. Maldacena, J.~Michelson, and A.~Strominger, ``Anti-de {S}itter
  fragmentation,'' {\em JHEP} {\bfseries 02} (1999) 011,
  \href{http://arXiv.org/abs/hep-th/9812073}{{\ttfamily hep-th/9812073}}.

\end{thebibliography}
\providecommand{\href}[2]{#2}\begingroup\raggedright\endgroup

\end{document}